\documentclass[aps,pra,twocolumn,showpacs,preprintnumbers]{revtex4}

\usepackage{epsfig}
\usepackage{CJK}
\usepackage{bm}
\usepackage{amssymb}
\usepackage{amsmath}
\usepackage{graphicx}
\begin{document}

\begin{CJK}{UTF8}{gkai}

\bibliographystyle{unsrt}


\title{\bf Spectral characterization of third-order harmonic generation assisted by two-dimensional plasma grating in air}
\author{P.J. Ding, Z.Y. Liu, Y.C. Shi, S.H. Sun, X.L. Liu, X.SH. Wang, Z.Q. Guo, Q.C. Liu, Y.H. Li, B.T. Hu}
\email[Email:]{hubt@lzu.edu.cn}
\affiliation{School of Nuclear Science and Technology, Lanzhou University, 730000, China}

\date{\today}

\begin{abstract}
A dramatic spectral modulation of third-order harmonic (TH) emission generated in a near infrared femtosecond (fs) pulse filamentation, assisted by a two-dimensional plasma grating formed by two pump femtosecond pulses, is experimentally demonstrated when their spatiotemporal overlap is achieved. It is mainly attributed to strong cross-phase modulation induced by the fundamental wave of the probe pulse and two pump ones. The delay dynamic of TH spectra indicates the influence of two retarded nonlinear responses on the TH generation. The dependences of TH generation on the energies of probe and pump pulses, relative field polarization angle are also studied.
\end{abstract}
\smallskip
\maketitle

\section{INTRODUCTION}
The generation of ultrashort pulses in the ultraviolet (UV) regime has stimulated many research interests in recent decade because it provides massive applications in many fields, such as ultrafast spectroscopy of atoms and molecules, time-domain studies of biological processes, pulse self-compression through filamentation and white-light continuum generation \cite{Nature.419.803, J.Phys.Chem.A.104.5660, Nature.418.290, Ghotbi:11, Laser.Physics.22.1, PhysRevLett.89.143901, springerlink:10.1007/s00340-003-1199-2, springerlink:10.1007/s00340-004-1689-x}. Nonlinear frequency up-conversion, especially TH generated by infrared (IR) femtosecond pulse, is one of the most efficient approaches to obtain intense, ultrashort laser pulse in the UV spectral region \cite{Reiter:10}. The spectral broadening of TH pulses generated during femtosecond pulse filamentation in air or noble gases has been applied in the generation of ultrashort pulses in the UV. For instance, by using spectral broadening of TH pulses in argon, sub-20-fs UV pulses with more than 300 $\mu J$ energy at 268 nm have been generated \cite{Ghotbi:11}. The effects of various physical mechanisms on spectral modulation of TH pulses generated during femtosecond pulse filamentation were interpreted in terms of spatiotemporal self-phase modulation (SPM) and cross-phase modulation (XPM) originated from electronic Kerr nonlinearity, free-electron generation, stimulated Raman scattering and temporal self-steepening \cite{springerlink:10.1007/s00340-003-1214-7}. The XPM originated from molecular alignment can further modulate the TH spectrum \cite{PhysRevLett.87.153902, PhysRevLett.100.123006}. However, complicated couplings of nonlinear effects during femtosecond laser filamentation caused by the interplays and competitions of these different nonlinear effects make the intrinsic physical mechanisms of filaments interaction difficult to be fully understood.

The TH generated in a probe filament could be dramatically enhanced through the nonlinear filament interaction with a pump filament or plasma grating, which closely depended on the relative polarizations, crossing angles, and intensity ratios of the probe and pump pulses \cite{yang:071108, Liu:12}. Since the inhomogeneous distribution of electron density in plasma grating, the probe pulse experiences fluctuation of refractive index in different positions, which should influence the generation and propagation dynamics of TH pulse. In this work, we experimentally investigated the influences of different interacting positions between a probe filament and a plasma grating generated by two non-collinear pump filaments on the generation and spectral modulation of TH pulses produced in the probe filament. An enhanced spectral modulation of TH is observed when the probe filament interacts with the forepart of the plasma grating. The delay dynamic of TH spectra is also studied to further understand the influence of other nonlinear effect on the generation of TH pulses. Energy exchange between the probe pulse and two pump ones when they experience filamentation process is observed, which indicates the influence of retarded rotational Raman effect from $O_{2}$ and $N_{2}$. And then, the dependences of TH generation on the energies of probe pulse and two pump ones, the polarization angle between the probe and pump pulses are investigated. 

This paper is organized as follows. In Section \ref{experi-setup} we describe our experimental setup. Results and discussion are presented in Section \ref{results-discussion} and we conclude in Section \ref{summary}.
\section{Experimental Setup}
\label{experi-setup}
The experimental setup is schematically illustrated in Fig. 1. The laser system is a Ti:Sapphire chirped-pulse amplifier system (Quantronix) capable of producing 33 fs pulses centered at 810 nm with 70 nm spectral bandwidth, at a repetition rate of 1 kHz. The transverse profile of intensity distribution is Gaussian-shaped, and the spot size at $1/e^2$ of the laser beam is about 10 mm. As shown in Fig.1, the output laser beam is split into three beams by two successive beam-splitters (BS1, BS2). One beam with pulse energy of 0.6 mJ, which is used as the probe beam, is focused by the lens L3 (f = 117 cm), leading to a 3 cm filament. The other two beams used as the pump beams named by Pump1 and Pump2, with pulse energy of 0.9 mJ and 0.7 mJ respectively, are focused by the lens L1 and L2 (f = 60 cm). The time delay between the three pulses is controlled by two motorized translation stages (Delay1, Delay2) with a precision of 2.5 $\mu m$. The two pump beams are spatiotemporally overlapped by adjusting the delay line Delay2 in order to generate a one-dimensional plasma grating with 2.5 cm length, and then by adjusting the delay line Delay1, the probe beam would be spatiotemporally overlapped with the two pump beams leading to a two-dimensional plasma grating in the core of interaction region as shown by the inset in Fig. 1. The crossing angle between the two pump beams marked as $\theta $ is fixed to 2.9 degree, and the crossing angle between the probe beam and the Pump2 marked as $\phi $ is fixed to 3.6 degree by three 800 nm high reflective mirrors M8, M9 and M11. The pulse energy of the probe or pump beam could be adjusted by inserting several attenuate plates. 

In order to obtain an optimal TH signal, the probe beam after filamentation interactions is collected onto a spectroscopy (Ocean Optics, USB4000) by the lens L4 after reflection by a 266 nm high reflective mirrors with spectral bandwidth of 63.8 nm. Several attenuation plates are placed before the entrance of the spectroscopy to avoid the saturation of the spectroscopy. A sensitive CCD (1024×1024 pixels, 2.2 $\mu m$) is installed on the top of the filament intersecting region to record the fluorescence image of the plasma grating, which is collected by using an Al-coated concave mirror installed on the bottom.

\begin{figure}[ht]
   \begin{center}
     \includegraphics[width=3.5in,angle=0,clip=true]{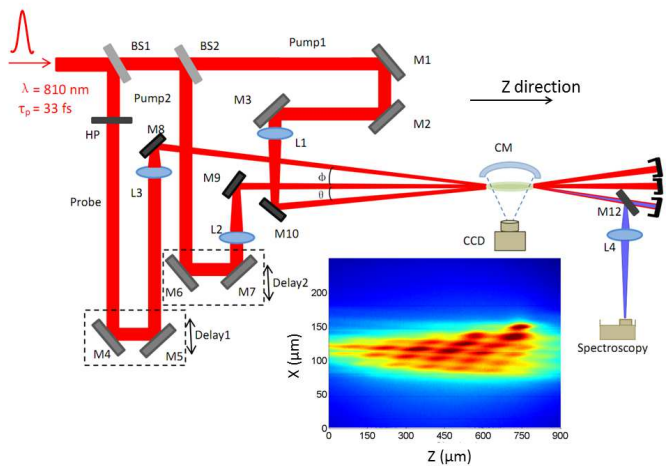}
     \caption{(Color online) Schematic illustration of the experimental setup for measuring the TH spectra and plasma fluorescence distribution of filaments interaction region. Here: BS1 and BS2, beam-splitter; L1, L2, L3, L4, focusing lenses; M1, M2, M3, M4, M5, M6, M7, M8, M9, M10, 800 nm high reflective mirrors; M11, 266 nm high reflective mirrors; CM，Al-coated concave mirror. The inset represents the two-dimensional plasma grating generated by the probe pulse and two pump pulses when they achieve spatiotemporal overlap.}
    \label{fig-1}
    \end{center}
\end{figure}

\section{Results and Discussion}
\label{results-discussion}
Intense interaction between three non-collinear filaments is observed as they are spatially overlapped and temporally synchronized. The density of plasma generated in the interaction region is modulated due to the constructive and destructive interference between three pulses with same polarizations. Local refractive index is changed due to the optical Kerr effect and creation of a periodic wavelength-scale plasma microstructure named by plasma grating \cite{yang:071108, Liu:12, PhysRevLett.107.095004, Yang:09} with dimensions of 0.90 mm $\times$ 0.75 mm, which is shown in the inset of Fig. 1. When the probe filament interacts with the centre of plasma grating marked by Position B as shown in Fig. 2, an extraordinary bright sparking spot accompanied by a blast sound is observed therein when the delay time is adjusted to an appropriate value. This strong fluorescence is considered as the sign of temporal synchronization and the corresponding delay time is defined as zero delay time. It evidently indicates a significant plasma generation and an increase of the local electron density. It can be contributed to the reason: the preformed electrons inside the one-dimensional plasma grating induced by two pump filaments are accelerated by the probe filament, leading to further inelastic collisional ionization \cite{PhysRevLett.107.095004}. 

In order to investigate the influence of plasma grating on the generation of TH pulse, the spectra of TH are recorded as the positions of interaction region between the probe filament and plasma grating varied after temporal synchronization, as shown in Fig. 2. The spectrum marked by Position C, which corresponds to the situation where the probe filament interacts with the rear-part of plasma grating with about 500 $\mu$m away from its centre, shows a symmetrical modulation while the central wavelength shifts to the red side a little. When the probe filament interacts with the forepart of plasma grating with about 300 $\mu$m away from its centre, the TH spectrum marked by Position A is largely modulated to blue side accompanied by a little redshift. Meanwhile, the generation of TH is significantly enhanced compared to that of Position B and Position C.

\begin{figure}[ht]
   \begin{center}
     \includegraphics[width=3.5in,angle=0,clip=true]{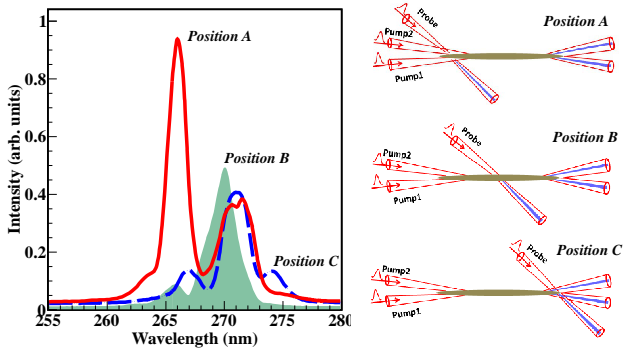}
     \caption{(Color online) The measured spectra of TH generated in the probe filament relative to the plasma grating for three different intersection positions, e.g. before the centre (Position A), in the centre (Position B) and behind the centre (Position C) of plasma grating. The filled curve shows the spectrum of TH in the case of Position B for comparison. The insets on the right are the corresponding schematic illustrations of filaments interaction positions.}
    \label{fig-2}
    \end{center}
\end{figure}

The spectral modulation caused by TH itself can be neglected because the intensities during the TH pulse filamentation are not high enough to cause strong SPM because of much lower clamped peak intensity compared to the corresponding intensity in the filament of IR fundamental pulse. Therefore, the change of refractive index experirenced by TH pulse is mainly due to the influence of XPM effect from fundamental wave (FW) of the probe and pump pulses and plasma formation. Concerning the influence of XPM effect, the shift of TH frequency can be expressed by \cite{Laser.Physics.22.1, Liu:11, wang:031105}
\begin{eqnarray}
\begin{aligned}
  \delta \omega_{TH}(t)=-\frac{3\omega_{0}}{c}z\frac{\partial }{\partial  t}[n_{2}I_{probe}(t) \\
   +n_{2}I_{pumps}(t)-\frac{5\omega_{p}^2(t)}{9\omega_{0}^2}]
\end{aligned}
\end{eqnarray}
where $I_{TH}$, $I_{probe}$ and $I_{pumps}$ respectively represent the intensity of TH pulse generated in the probe filament, the probe pulse and total laser field coupled by two pump pulses, $z$ is the interaction length of filaments, $n_{2}$ is the Kerr nonlinear coefficient originated from XPM effect, $\omega _{0}$ is the angular frequency of IR fundamental pulses and $\omega _{p}(t)= [4\pi e^2 N_{e}(t)/m_{e}]^{1/2}$ is the critical plasma frequency where $N_{e}$ is the electron density. 

Equation (1) clearly shows that the plasma effect shifts the frequency of TH only to the blue side due to the negative contribution to the refractive index from the free electrons while the XPM effect originated from the temporal variation of field intensity of the probe and pump pulses can cause spectral modulation of TH to both sides. In the case of Position B as shown in Fig. 2, further inelastic collisional ionization triggered by the probe filament considerably enhances the generation of free electrons, which confines the red-shift and only leads to a limited blue-shift of TH spectrum. But in the cases of Position A and Position C, the departures from spatial overlap make the generation of local electron plasma not enough to confine spectral red-shift, so their corresponding TH spectra show two-sides modulation. In addition, the TH spectrum in the case of Position A shows a larger blue-shift of frequency components compared to that in the case of Position B. In this case, the probe filament interacts with the forepart of plasma grating, where the strong XPM by FW of the probe and two pump pulses dominates the spectral modulation of TH pulses because the influence of plasma genration could be weak. On the other hand, in the case of Position C where the probe filament interacts with the rear-part of plasma grating, the energy of pump filament has slightly decreased due to the ionization process during filamentation. In addition, the pumps start to defocus at the rear-part of plasma grating resulting to a decreased field intensity. Correspondingly, the intensity of two pumps will decrease, leading to the weakening of XPM effect. Thus, the spectral modulation of TH is relatively weaker in the case of Position C compared to that in the case of Position A.

In order to further investigate the spectral characteristics of TH pulses generated in the probe filament, the delay dynamic of TH generation in the case of Position B is studied by tuning the time delay between the probe and two pump beams, as shown in Fig. 3(a). The positive delay refers to the probe pulse propagating behind two pump pulses. Compared to that of zero delay, TH exhibits a larger generation when the probe pulse propagates behind two pump pulses with 33.4 fs, which approximately equals the duration of input laser pulse. On the contrary, the propagation of the probe pulses before the pump pulses with 33.4 fs, in which case where the pump and probe pulse are temporally synchronized almost within the pulse duration, leads to a smaller TH spectral intensity compared to that of zero delay. When the delay is varied from 0.05 ps to 0.10 ps, the spectral modulation of TH pulses to blue side is evident while the spectral intensity is reduced as shown in Fig. 3(b). The enhancement of TH generation is still observed for 0.3 ps after the extinction of two pump pulses. 

\begin{figure}[ht]
   \begin{center}
     \includegraphics[width=3.5in,angle=0,clip=true]{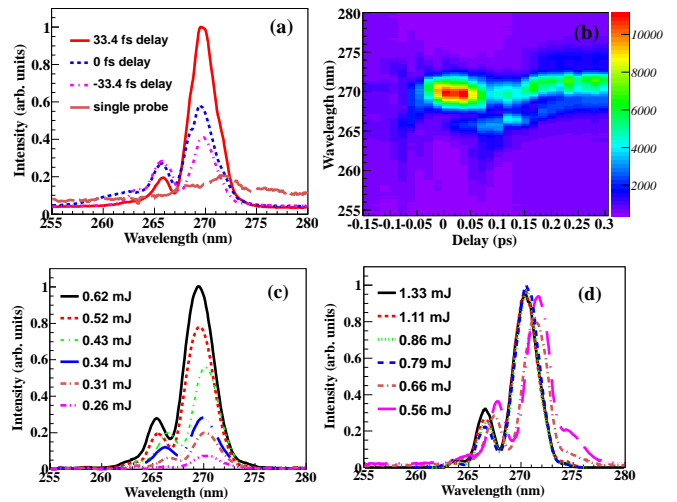}
    \caption{(Color online) (a) TH spectra recorded at three different conditions where time delays between the probe and two pump pulses are 0 fs, 33.4 fs and $-33.4$ fs respectively. The spectrum of TH generated in the probe filament without two pump pulses, which is enlarged by 200 times, is also recorded for comparison. (b) TH spectra as a function of time delay between probe pulse and two pump pulses. [(c)-(d)] TH spectra generated in probe filament as a function of the probe energy and pump energies respectively when three beams are tuned to match spatiotemporal overlap.}
    \label{fig-3}
    \end{center}
\end{figure}

In general, the nonlinear response of a transparent medium to an intense and ultrashort laser field consists of an instantaneous and a retarded change of the refractive index, so the $n_{2}$ term in equation above also consists two corresponding parts. Unlike the decaying exponential usually assumed as the temporal response in nonlinear optics, air has a more complicated temporal response function $R(t)$ which could be calculated from quantum mechanical consideration of $O_{2}$ and $N_{2}$ molecular rotation \cite{Ripoche1997310}. Then we have the overall Kerr nonlinear response:
\begin{eqnarray}
  n_{2}I(t) = n_{2,i}I(t) + n_{2,d}\int _{-\infty }^{t} R(t^{'})I(t-t^{'})dt^{'},
\end{eqnarray}
where $n_{2,i}$ is the instantaneous electronic nonlinear coefficient, $n_{2,d}$ is the delayed nonlinear coefficient, $I(t)$ is the intensity of the probe or pump pulses. It has been determined by J. F. Ripoche  $et$ $al$ that $n_{2,i}\approx n_{2,d}$ \cite{Ripoche1997310}. For the retarded refractive index change originated from plasma generation process, it is directly proportional to the plasma density $N_{e}(t)$, which can be estimated by $N_{e}(t)=\int _{-\infty}^{t} \sigma _{8} \rho _{mol} I^{8}(t^{'}) dt^{'}$ \cite{Couairon200747}. The terms $\sigma _{8}$ and $\rho _{mol}$ are the cross section of multiphoton ionization and the density of neutral molecules. 

The change of local refractive index is originated from two retarded nonlinear responses in air, e.g. the rotational Raman effect and the formation of a plasma grating \cite{PhysRevLett.102.123902, PhysRevLett.105.055003}. For the retarded Raman response of the molecule characterized by the temporal response function $R(t)$ in equation (2), the corresponding temporal retardation with respect to the zero delay is correlated with the characteristic time of molecule response to the laser field, which can be estimated as tens of fs for a pulse at 800 nm in air \cite{PhysRevLett.105.055003}. It has been realized that only the retarded nonlinear response of the molecule can result in energy transfer between two filaments because it induces a spatial phase shift between the laser intensity interference and the refractive index modulation. In turn the existence of the retarded rotational Raman effect can be confirmed by the presence of energy exchange between filaments. Figure 4 shows results obtained when the energy of probe pulse is attenuated down to 0.546 mJ and the average energy of two pump pulses is also attenuated down to 0.635 mJ. We observe an energy transfer between the probe filament and two pump filaments. There are two different regimes. These results are similar to the observation by Bernstein et al \cite{PhysRevLett.102.123902}, which indicates the effect of retarded rotational Raman effect. For the retarded nonlinear response originated from the formation of two-dimensional plasma grating, the plasma grating stays at rest at the end of two pump pulses overlap for a long time after its production process ($\sim$120 fs) \cite{PhysRevLett.105.055003} because its decay time is much larger than the time scale of pulse duration. 

\begin{figure}[ht]
   \begin{center}
     \includegraphics[width=3.4in,angle=0,clip=true]{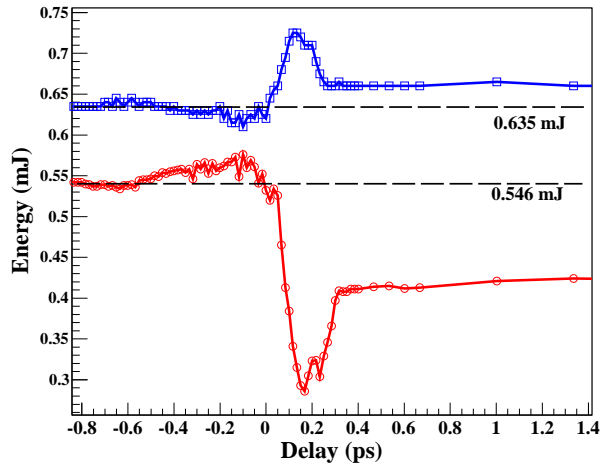}
    \end{center}
    \caption{(Color online) Energy exchange of between the probe pulses (red circle) and two pump pulses (blue rectangle) in air as a function of time delay.}
    \label{fig-4}
\end{figure}

Three-order of magnitude enhancement of TH generation with respect to that of single filament can be observed as shown in Fig. 3(a). It is originated from the combination of the plasma-enhanced bulk third-order optical susceptibility $\chi ^{(3)}$ due to plasma formation and the breakdown of Gouy phase shift. The underlying mechanism of first effect is that the presence of free electrons and ions can effectively increase the third-order nonlinear response of a medium. As is well known, the Gouy phase shift effect can result in a large cancellation of TH generation \cite{Rodriguez:11}. The presence of the intercepting plasma grating could break it, leading to a large enhancement of TH generation with respect to that in the case of single filament. The relative enhancement of TH generation in delay time of 33.4 fs compared to zero delay time in Fig. 3(a) could be ascribed to the retarded impulsive rotational Raman effect. The observed subsequent spectral intensity reduction in Fig. 3(b) at higher delays is due to the plasma density decay in the plasma grating, which is similar to the observation in \cite{PhysRevA.81.033817}. One can expect that the TH generation gradually decrease with the increasing delay time.  

The output spectra of TH pulses as a function of various input probe energies and pump energies are also investigated as shown in Fig. 3(c) and Fig. 3(d). The pulse energy of the incident probe beam is changed from 0.26 mJ to 0.62 mJ, and the TH spectra are measured when the time delay set to 0 fs in the case of Position B. The generation of TH is reduced due to the weakening of field intensity inside the probe filament with the decrease of input probe energies. This result can be easily understood because the intensity of TH is proportional to the third power of the probe energy. However, in the presence of the spatiotemporal overlapped plasma grating, the generation and spectral modulation of TH pulses weakly depend on the pump energies when it varies from 0.56 mJ to 1.33 mJ, indicating that the observed enhancement of TH generation in the case of Position B is originated from the interception of plasma grating and has no close relation with the pump energies. This behavior is correlated with the saturation of plasma density. At high plasma density, the TH generation will saturate as a consequence of the independence on plasma density \cite{feng:072305, PhysRevA.81.033817}. This result clearly gets rid of the influence of XPM effect induced by two pump pulses in the case of Position B.

\begin{figure}[ht]
   \begin{center}
     \includegraphics[width=3.5in,angle=0,clip=true]{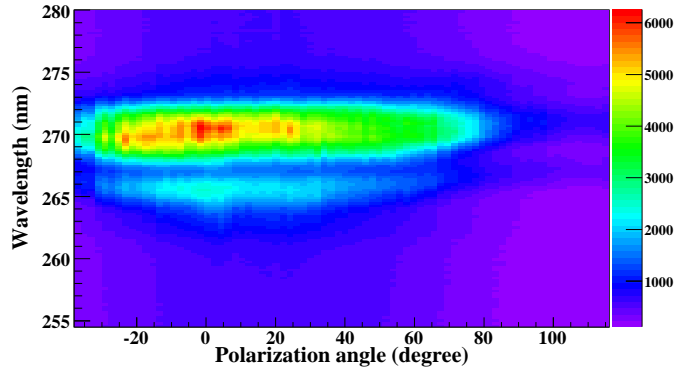}
    \end{center}
    \caption{(Color online) Recorded TH spectra as a function of varying field polarization angle between the probe pulse and the two pump pulses in air.}
    \label{fig-5}
\end{figure}

We then investigated the spectra of TH induced in the probe filamentation by varying the field polarization of the probe pulse with respect to the two pump pulses in the situation where spatiotemporal overlap is achieved, as shown in Fig. 5. The field polarization angle, at which the probe pulse parallels with plasma grating characterized by the strongest plasma fluorescence, is defined as zero field polarization. Obviously, the TH generation along the probe beam gets the maximum or minimum as the probe polarization is parallel or orthogonal to the pump polarization respectively, corresponding to the constructive and destructive interference between the probe and pump pulses. So the transform from constructive interference to destructive interference results in a weakening of instantaneous XPM effect when the relative field polarization angle varies from zero degree to 90 degree, which leads to a gradual decay of the whole TH generation down to the minimum. In addition, the $n_{2}$ originated from two pump pulses in the case of parallel polarization is three times larger than that in the case of orthogonal polarization \cite{Bejot:08}.

\section{Summary}
\label{summary}
In summary, a dramatic spectral modulation of TH generated in a probe pulse filamentation is observed when the probe pulse interacted with the forepart of plasma grating generated by two non-collinear pump filaments. It is mainly attributed to strong XPM effect induced by the FW of the probe pulse and two pump ones. The investigation of TH delay dynamics indicates the influence of two retarded nonlinear responses on the generation of TH, e.g. the rotational Raman effect and the formation of plasma grating. The existence of former has been confirmed by the energy exchange between the probe filament and two pump filaments. The independence of TH intensity on the energies of two pump pulses indicates the saturation of plasma density. The spectral intensity of TH shows a maximum and minimum for the parallel and orthogonal polarization between the probe and two pump pulses respectively. These results are promising for several important applications in studying the generation of harmonic emission and frequency up-conversion leading to ultrashort laser pulse in deep UV regime in gases.

\section{Acknowledgments}
  We gratefully acknowledge the support of National Natural Science Foundation of China (Grant No.11135002, 11075069, 91026021, 11075068 and 10975065), and the Fundamental Research Funds for the Central Universities (Grant No. lzujbky-2010-k08).
\bibliography{paper_submitted_to_pra}

\end{CJK}
\end{document}